\documentclass[
reprint,
superscriptaddress,
nofootinbib,
amsmath,amssymb,
aps,
prd,
floatfix,
]{revtex4-2}
\usepackage{amsmath, amssymb}
\usepackage{listings}
\usepackage{graphicx}
\usepackage{dcolumn}
\usepackage{bm}
\usepackage{float,color}
\usepackage[normalem]{ulem}
\usepackage{tabularx}
\usepackage{mathtools} 
\usepackage{multirow}

\usepackage{scrextend}
\usepackage{geometry}
\usepackage{scrextend}
\usepackage{hyperref}

\begin{document}

\title{Octupolar test of general relativity
}

\author{Parthapratim Mahapatra}
\email{ppmp75@cmi.ac.in}
\affiliation{Chennai Mathematical Institute, Siruseri, 603103, India}
\date{\today}

\begin{abstract}
Compact binaries with unequal masses and whose orbits are not aligned with the observer's line of sight are excellent probes of gravitational radiation beyond the quadrupole approximation. Among the compact binaries observed so far, strong evidence of octupolar modes is seen in GW190412 and GW190814, two binary black holes observed during the first half of the third observing run of LIGO/Virgo observatories. These two events, therefore, provide a unique opportunity to test the consistency of the octupolar modes with the predictions of general relativity (GR). In the post-Newtonian (PN) approximation to GR, the gravitational-wave phasing has known dependencies on different radiative multipole moments, including the mass octupole.
This permits the use of publicly released posteriors of the PN phase deformation parameters for placing constraints on the deformations to the different PN components of the radiative mass octupole denoted by $\delta \mu_{3n}$.
Combining the posteriors on $\delta \mu_{3n}$ from these two events, we deduce a joint bound (at 90\% credibility) on the first three PN order terms in the radiative octupoles to be $\delta \mu_{30}=-0.07^{+0.11}_{-0.12}$, $\delta \mu_{32}=0.48^{+0.93}_{-1.15}$, and $\delta \mu_{33}=-0.32^{+1.67}_{-0.62}$, consistent with GR predictions. Among these, the measurement of $\delta \mu_{33}$ for the first time confirms the well-known octupolar tail contribution, a novel nonlinear effect due to the scattering of the octupolar radiation by the background spacetime, is consistent with the predictions of GR. Detection of similar systems in the future observing runs should further tighten these constraints.
\end{abstract}


\maketitle
\section{Introduction}\label{sec:intro}
It is well known that the leading order gravitational wave (GW) emission is \emph{quadrupolar} according to general relativity.
However, subdominant \emph{higher multipoles} get turned on if the binary has a \emph{mass asymmetry} and when the line of sight of the observer is \emph{not aligned} with the orbital angular momentum vector of the binary \cite{BFIS08,VanDenBroeck:2006qu,VanDenBroeck:2006ar, ABFO08, MKAF2016, IMRPhenomHM, Roy:2019phx, Khan:2019kot, Cotesta:2018fcv}. 
To date, the LIGO-Virgo-KAGRA collaboration has reported $\sim$90 confident detections of compact binary mergers \cite{GW150914,GWTC1,GWTC2,GWTC3}. Among these events, two compact binary mergers---GW190412 \cite{GW190412} and GW190814 \cite{GW190814}---have shown clear evidence of the presence of \emph{octupolar} ($\ell=3, m=3$) mode, the first correction beyond the quadrupole. These two events, therefore, should facilitate a test of the gravitational octupolar structure of the compact binary dynamics.

The gravitational dynamics of a compact binary system is typically divided into three stages of evolution: inspiral, merger, and ringdown. While the post-Newtonian (PN) approximation to  general relativity (GR) \cite{Bliving} is employed to model the adiabatic inspiral stage of a compact binary coalescence, one requires numerical solutions to the Einstein equations \cite{Pretorius07Review}, and the black hole (BH) perturbation theory \cite{TSLivRev03} to describe the highly nonlinear merger stage, and the ringdown phase, respectively. As numerical relativity simulations are computationally expensive, currently there are two main modeling approaches towards producing the complete gravitational waveform (i.e., a single waveform that captures all three stages of binary evolution) for parameter inference: effective one-body approach \cite{Damour-EOB, Buonanno-EOB} and phenomenological approach \cite{AjithNR07b,Ajith09}. Both these methods make the best use of the analytical and numerical understanding of compact binary dynamics.

The gravitational waveform from a coalescing compact binary within GR, in the frequency domain, has the following form:
\begin{equation}\label{eq:waveform}
    \Tilde{h}(f; \vec{\lambda}, \iota,  \varphi_{N})= \sum_{\ell \ge 2} \sum_{m=-\ell}^{\ell} \, Y_{-2}^{\ell m}(\iota,  \varphi_{N}) \, \Tilde{h}_{\ell m} \, (f;\vec{\lambda}) ,
\end{equation}
where $Y_{-2}^{\ell m}$ are spin-weighted spherical harmonics of spin weight $-2$, ($\iota, \varphi_{N}$) describes the location of the observer in the binary's sky and $\vec{\lambda}$ denotes the intrinsic parameters (e.g., masses and spins) as well as other relevant extrinsic parameters (e.g., luminosity distance ($d_{L}$), reference time and reference phase) of the binary. Each GW mode ($\Tilde{h}_{\ell m}$) has an amplitude, $A_{\ell m}(f;\vec{\lambda})$, and a phase, $\psi_{\ell m}(f;\vec{\lambda})$ (i.e., $\Tilde{h}_{\ell m}=A_{\ell m}(f;\vec{\lambda}) \, e^{i \, \psi_{\ell m}(f;\vec{\lambda})}$). In alternative theories of gravity, the gravitational dynamics of a compact binary could differ from the prediction of GR during all the three stages and might modify the phase and amplitude in the waveform.

There exist proposals in the literature to probe the prediction of the harmonic structure of gravitational radiation from binary black hole coalescence in GR ~\cite{Dhanpal:2018ufk, Islam:2019dmk, Kastha:2018bcr, Kastha:2019, Mezzasoma:2022pjb}. Using GW190412 and GW190814, Ref.~\cite{Capano:2020dix} tested the consistency between the dominant and subdominant modes and found the chirp mass estimated from the $\ell=3,m=3$ mode to be within $\pm 1\%$ of the one estimated from the quadrupolar $\ell=2, m=2$ mode.

In a more recent work~\cite{Puecher:2022sfm}, the consistency of the amplitudes of the $h_{21}$ and $h_{33}$ modes of the GW spectrum with GR predictions was investigated using these two events and found no evidence for any violation of GR. This test assumes the phases of subdominant harmonics ($\psi_{\ell m;\ell>2}$) follow GR and investigates whether the amplitudes of subdominant harmonics ($A_{\ell m;\ell>2}$) are consistent with the GR prediction. 

In this paper, we argue that if a signal contains nonquadrupolar modes, apart from the amplitude, the phase evolution will also carry their unique imprints~\cite{BDI95,BFIS08,Kastha:2018bcr,Kastha:2019}. As GW detectors are more sensitive to \emph{phase evolution}, this could be used to test GR, complementing the approach of \cite{Puecher:2022sfm}. However, we will focus only on the inspiral phase in this work, which is well-modeled by PN approximation to GR, and discuss constraints on the \emph{PN structure} of octupolar emission in GR. For this, we will make use of the unique map between the mass-type octupole coefficients at different PN orders and the bounds on the 1PN, 2PN, and 2.5PN logarithmic phasing deformation parameters for these two events in the \emph{parametrized} tests of GW phasing reported in \cite{TGR-GWTC-2,TGR-GWTC-3}. Further, we will consider only the leading order appearance of the octupole coefficients in the GW phase for this mapping.

The remainder of the paper is organized as follows. In Sec.~\ref{sec:parametrized-tests}, we briefly review the parametrized tests of GW phasing. In Sec.~\ref{sec:multipolar-waveform}, we introduce the octupolar parametrization. We derive the relations between different PN pieces in the mass-type octupole moment and different PN phasing terms in Sec.~\ref{sec:mapping}. In Sec.~\ref{sec:inference}, we describe the Bayesian framework to infer the octupole parameters. Our results and conclusions are presented in Sec.~\ref{sec:results}.

\section{Parametrized tests of GW phasing}\label{sec:parametrized-tests}
The frequency domain GW phase from the inspiral part of the waveform (computed using the stationary phase approximation \cite{Cutler:1994ys,DIS00}) for the leading quadrupolar harmonic \cite{BIOPS2009,ABFO08} takes the form
\begin{align}
\Psi(f)&=2\pi f\,t_c-\phi_c \nonumber \\
&\quad+\frac{3}{128\,\nu\,v^5}\left[ \, \sum_{i=0}^{i=7} \left( \phi_i \, + \phi_{il} \, \ln v \right) v^i \, + \mathcal{O}(v^{8}) \right] \,,
\end{align}
where $v=(\pi G M f/c^{3})^{1/3}$ is the PN expansion parameter,  $M$ is the binary's redshifted total mass, $\nu$ is the symmetric mass ratio of the binary, $\phi_{i}$ and $\phi_{il}$ denote the nonlogarithmic and logarithmic PN phasing coefficients, respectively.

Due to the lack of accurate waveforms in alternative theories of gravity, ``theory-agnostic" approaches are often adopted to test GR with GW data.  These ``null tests" of GR make use of our best knowledge of compact binary dynamics in GR and look for possible deviations from GR without reference to specific alternatives (see Refs.~\cite{TGR-GW150914,TGR-GWTC-1,TGR-GWTC-2,TGR-GWTC-3} for more details.). One of the most generic tests of GR that has been routinely performed with LIGO/Virgo data is the parametrized test of GW phasing \cite{BSat94,BSat95,AIQS06a, AIQS06b,YunesPretorius09, MAIS10,Li:2011cg,TIGER,Mehta:2022pcn}. 

The parametrized tests rely on measuring any deviations in the PN coefficients $\phi_i$ and $\phi_{il}$ in the GW phasing, which are uniquely predicted by GR, from compact binary mergers. A parametrized waveform model introduces additional degrees of freedom to capture signatures of possible GR violation by modifying the phasing coefficients as
\begin{equation}
    \phi_b=\phi_b^{\rm GR}(1+\delta \hat{\phi}_b),\label{eq:param}
\end{equation}
($b=i,\,il$) (see Sec. VA of Refs.~\cite{TGR-GWTC-2,TGR-GWTC-3} for more details). 
In GR, these phenomenological \emph{dimensionless deviation} parameters ($\delta \hat{\phi}_b$) are identically zero, whereas in alternative theories of gravity, one or more of these parameters could be different from zero. Combining data from different GW events detected during the first, second, and third observing runs of LIGO/Virgo, the current bound on the PN deviation parameters are found to be consistent with GR (see Figs. 6 and 7 of Ref. \cite{TGR-GWTC-3}).

For the two asymmetric binary events, GW190412 and GW190814, we will use the results of the parametrized tests, obtained by applying parametrized {\tt IMRPhenomPv3HM} (denoted as ``{\tt Phenom}" in this paper) \cite{Khan:2019kot} and {\tt SEOBNRv4HM$\_$ROM} (denoted as ``{\tt SEOB}")\cite{Cotesta:2018fcv} waveform approximant to the data. {\tt Phenom} waveform is a frequency-domain phenomenological waveform model that includes the effects of two-spin precession along with higher multipole moments \cite{Khan:2019kot}, whereas {\tt SEOB} is a frequency-domain nonprecessing reduced-order effective one-body model that incorporates the higher order modes \cite{Cotesta:2018fcv}.
In the current LIGO-Virgo-KAGRA analyses \cite{TGR-GWTC-1,TGR-GWTC-2,TGR-GWTC-3}, the reported bounds on $\delta \hat{\phi}_{b}$ come from the fractional deviations applied to the nonspinning portion of the phase (see Sec.~VA of Refs.~\cite{TGR-GWTC-2,TGR-GWTC-3} for detailed discussions).

\section{Parametrized multipolar gravitational waveforms}\label{sec:multipolar-waveform}
The \emph{radiative} multipole moments of a compact binary system contain information about source physics (masses \cite{BD92, BDI95,Bliving}, spins \cite{KWWi93,Apos95,K95,Poisson97, BBuF06,ABFO08,BBF2011,BMFB2012, BMB2013, Bliving, MBFB2012,M3B2013, Marsat2014, BFH2012,WWi96}, tidal deformability \cite{BC92ApJ,KC92ApJ,Lai94,KS95,MW04,FH08,HFB2020}, spin induced quadrupole moment \cite{Poisson97,Ryan97,Laarakkers99ApJ,Pappas:2012ns,Uchikata:2015yma} etc.) and account for various nonlinear interactions and physical effects (such as ‘‘tail’’ effects \cite{BD88,BS93,BDI95}, tails of tails~\cite{B98tail}, tail square~\cite{B98quad}, memory effects \cite{Chr91,Th92,ABIQ04,Favata08}, spin-orbit effects \cite{KWWi93, Blanchet:2006gy}, spin-spin effects \cite{KWWi93,BFH2012} etc.) that occur at different PN orders in GR. In alternative theories of gravity, one or more radiative multipole moments of a compact binary could be different from those in GR (see for instance Refs.~\cite{Endlich:2017tqa,Battista:2021rlh,Battista:2022hmv}). One can put constraints on such theories by studying the multipolar structure of asymmetric compact binary systems like GW190412 and GW190814.

References~\cite{Kastha:2018bcr,Kastha:2019} came up with a novel theory-agnostic method to test the multipolar structure of the gravitational field radiated from an inspiralling compact binary. Using the multipolar post-Minkowskian formalism \cite{Th80,BD84,BD86,B87,BD88,BD89,BD92,B95,BDI95,BIJ02,DJSdim,BDEI04}, Ref.~\cite{Kastha:2018bcr} derived the \emph{parametrized multipolar} gravitational wave phasing up to 3.5PN order for nonspinning binaries and Ref.~\cite{Kastha:2019} extended it for nonprecessing binaries.
The multipolar post-Minkowskian formalism relates the radiation content in the far zone, encoded in the mass- and current-type radiative multipole moments $\{U_L,V_L\}$, to the stress-energy tensor of the source. In order to model possible deviations in the multipole structure, Refs.~\cite{Kastha:2018bcr,Kastha:2019} adopted the following parametrization for radiative multipole moments: $U_L\rightarrow \mu_l\,U_L$, $V_L \rightarrow \epsilon_l \ V_L$. By construction, the phenomenological multipole parameters $\mu_l$, $\epsilon_l$ are equal to unity in GR. With this parametrization, the contributions from various radiative multipole moments to the GW phasing can be tracked separately, thereby facilitating tests of the multipolar structure of the PN approximation to GR.

In this work, we go one step further and probe the different PN orders in the radiative mass octupole moment of a compact binary as it evolves through the \emph{adiabatic inspiral} phase. We propose the parametrization
\begin{equation}\label{eq:parametrized-Uijk}
    U_{ijk} \longrightarrow \sum_{n} \frac{1}{c^{n}}\, \mu_{3n} \, U_{ijk}^{(n), {\rm GR}} \, ,
\end{equation}
where $U_{ijk}$ is the mass-type radiative octupole moment, $U_{ijk}^{(n), {\rm GR}}$ is the (n/2)th PN correction to $U_{ijk}$ in GR and $\mu_{3n}$ is the corresponding octupole coefficient. Note that there is no 1/c (i.e., 0.5PN) contribution in the mass octupole moment in GR. The 1.5PN correction term in the octupole moment arises due to the tail effect, caused by the scattering of the outgoing octupolar wave off the background spacetime associated with the total [Arnowitt-Deser-Misner (ADM)] mass of the source \cite{BD88,BS93,BDI95,B98tail,B98quad}. By definition $\mu_{3n}$ is unity in GR and appears at different PN orders in the phasing formula. For instance, $\mu_{30}$ first appears at 1PN, $\mu_{32}$ at 2PN and $\mu_{33}$ at 2.5PN (logarithmic) order. We next discuss how the existing bounds on the PN deformation coefficients $\delta \hat\phi_b$, based on the parametrization in Eq.~(\ref{eq:param}), reported in \cite{TGR-GWTC-2,TGR-GWTC-3} can be mapped to the bounds on $\mu_{3n}$ in the parametrization derived above.

\section{Mapping the PN bounds to the octupole parameters}\label{sec:mapping}
Each of the parameters $\mu_{3n}$ appears at multiple PN orders in the GW phasing. For example, $\mu_{30}$ appears at 1PN, 2PN, 2.5PN (logarithmic), and 3.5PN orders. Therefore if there is a deviation from GR in one of the $\mu_{3n}$, it will result in a dephasing of each of the PN phasing coefficients at the order in which this octupole parameter contributes.\footnote{We assume the corrections to the radiative mass-type quadrupole moment $\mu_{2n}$ follows GR. This is a reasonable assumption confirmed by the previous tests of GR performed on GW events of nearly equal mass systems for which all PN deformation parameters are fully functions of different $\mu_{2n}$.}
Here we neglect the modification to all PN orders except the leading order at which they first appear. This is a reasonable assumption to make because if there is a deviation in any of the $\mu_{3n}$, the leading order at which they appear would be most sensitive to such a deviation.

Therefore, the goal now will be to obtain constraints on $\mu_{30}$, $\mu_{32}$, and $\mu_{33}$ using the bounds on 1PN, 2PN, and 2.5PN logarithmic phase deformation parameters, respectively, along with other relevant intrinsic binary parameters for particular GW events. Further, while estimating bounds on one of the $\mu_{3n}$, we assume all other mass-type octupole parameters as well as rest of the multipole parameters to take their values in GR (i.e., $\mu_{3 n^{\prime};n^{\prime} \neq n}=\mu_{l;l\neq3}=\epsilon_{l}=1$) in the spirit of a \emph{single-parameter} test, i.e., varying one deformation parameter at a time.

The expression for 1PN phasing coefficient in the parametrized multipolar GW phase [see Eq. (2.16) of Ref. \cite{Kastha:2018bcr}] is given by
\begin{align} \label{eq:phi2}
   \phi_{2} & = \left(\frac{1510}{189}-\frac{130}{21}\nu\right)+ \bigg(\frac{\mu_{30}}{\mu_{2}} \bigg)^{2} \left(-\frac{6835}{2268}+\frac{6835}{567}\nu\right)
    \nonumber \\
    &\quad + \bigg(\frac{\epsilon_{2}}{\mu_{2}} \bigg)^{2} \left(-\frac{5}{81}+\frac{20}{81}\nu\right) \nonumber\\
    & =\phi_{2}^{\rm GR} -  K_1(\nu) \, \bigg[\bigg(\frac{\mu_{30}}{\mu_{2}} \bigg)^{2}-1\bigg]  \nonumber\\
    &\quad - K_2(\nu) \, \bigg[\bigg(\frac{\epsilon_{2}}{\mu_{2}} \bigg)^{2}-1\bigg], 
\end{align}
where
\begin{subequations}
\begin{align}\label{eq:1PNPhase}
\phi_{2}^{\rm GR}\ =\left(\frac{3715}{756} + \frac{55}{9} \nu\right) \, , \\
K_{1}(\nu)= \left(\frac{6835}{2268} - \frac{6835}{567}\nu\right) \, ,\\
K_{2}(\nu)= \left(\frac{5}{81} - \frac{20}{81}\nu\right) \ ,
\end{align}
\end{subequations}
and $\mu_2$ and $\epsilon_{2}$ are the mass and current quadrupole parameters, respectively.
  
Comparing the parametrized PN phasing of Eq.~(\ref{eq:param}) to Eq.~(\ref{eq:phi2}) we have the following relation between $\delta \hat{\phi}_2$ and $\mu_{30}$: 
\begin{equation}
    \delta \hat{\phi}_{2} = \frac{K_1(\nu) \, \bigg[1 - \Big(\frac{\mu_{30}}{\mu_{2}} \Big)^{2}\bigg] + \  \bigg[1 - \Big(\frac{\epsilon_{2}}{\mu_{2}} \Big)^{2}\bigg] \, K_{2}(\nu)}{\phi_{2}^{\rm GR}}.
\end{equation}
In the spirit of null tests, we find it more convenient to employ octupole deformation parameters $\delta \mu_{3n}$ defined for mass octupole by imposing $\mu_{3n} = 1 + \delta \mu_{3n}$ for the different PN pieces. The aim now would be to derive bounds on $\delta \mu_{30}$, $\delta \mu_{32}$, and $\delta \mu_{33}$  using GW observations assuming $\mu_{l;l\neq3}=\epsilon_{l}=1$ and $\delta \mu_{3 n^{\prime};n^{\prime}\neq n}=0$.\footnote{This would amount to assuming, for instance, $\delta\mu_{30}=0$ at all orders except at 1PN where it first appears. This is justified only if the posterior on $\delta\mu_{30}$ derived from 1PN is consistent with GR which is exactly what we find in this work. The same argument would apply to all higher-order $\delta \mu_{3n}$ that we deal with in this work.}

With the above assumption we can express $\delta\mu_{30}$ in terms of $\delta {\hat \phi}_2$ as
\begin{equation}\label{eq:bound_deltamu30}
\delta \mu_{30}=-1 \pm \sqrt{1 \ -  \ \delta \hat{\phi}_{2} \ \frac{\phi_{2}^{\rm GR}}{K_{1}(\nu)}}\, . 
\end{equation}
Among the two solutions, we will adopt the one which respects the GR limit (i.e., $\delta \mu_{30}$ vanishes when $\delta \hat{\phi}_{2} \ \rightarrow 0$).
Similarly, we also obtain the expressions for  $\delta \mu_{32}$ and $\delta \mu_{33}$ as (see Sec.~\ref{appendixoct} of the Supplemental Material for a derivation):
\begin{align}
  &\delta \mu_{32} = \frac{\phi_{4}^{\rm GR} \, \delta \hat{\phi}_{4}}{K_{3}(\nu)} \label{eq:bound_deltamu32} \, , \\ 
  &\delta \mu_{33} = \frac{\phi_{5l}^{\rm GR} \, \delta \hat{\phi}_{5l}}{K_{4}(\nu)}\label{eq:bound_deltamu33}\, .
\end{align}
The expressions of $\phi_{4}^{\rm GR}$, $K_{3}(\nu)$, $\phi_{5l}^{\rm GR}$, and $K_{4}(\nu)$ are provided in Sec.~\ref{appendixoct} of the Supplemental Material.

Having obtained the mapping, now the problem essentially amounts to using the posterior samples of \{$\delta {\hat \phi}_b, \nu$\} for any event and computing the corresponding posteriors on \{$\delta \mu_{3n}$\} using the above equations. The exact procedure followed is discussed next.

\begin{figure*}
\centering
\includegraphics[width=0.96\textwidth]{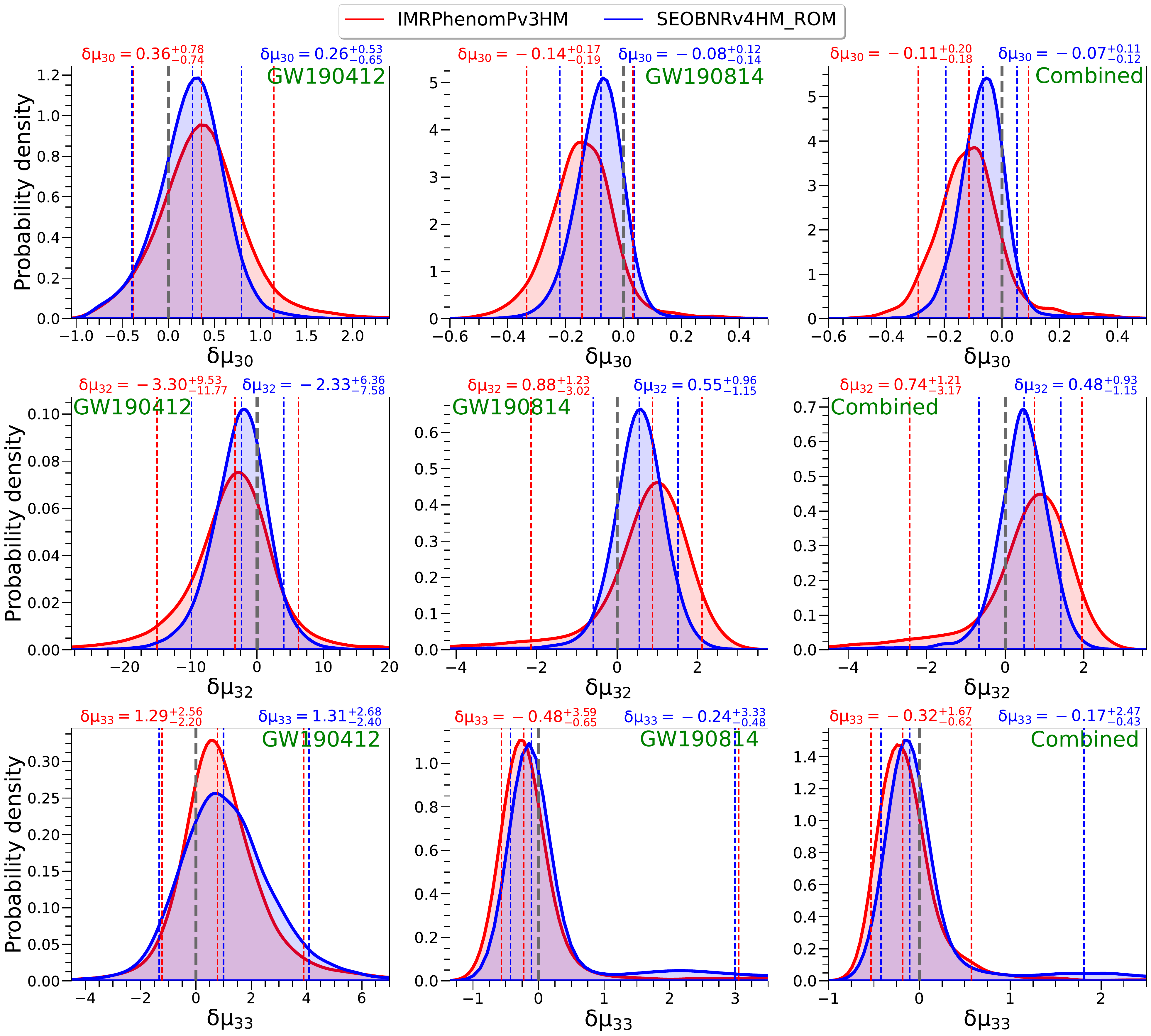}
\caption{\label{fig:dmu30-dmu32-dmu33-bounds} Bounds on $\delta \mu_{3n}$ for  GW190412 and GW190814 analyzed with {\tt Phenom} (in red color) and {\tt SEOB} (in blue color) waveform approximants are shown. Left-most panels show the bounds for GW190412, whereas the middle panels show the result for GW190814. The combined bounds are shown in the rightmost panels.
The colored vertical dashed lines mark the 90\% credible intervals and median values. The gray dashed vertical lines indicate the GR prediction ($\delta \mu_{3n}=0$). The posterior distributions of $\delta \mu_{3n}$ show consistency with GR.
}
\end{figure*}

\section{Inferring octupole parameters}\label{sec:inference}
Given the LIGO/Virgo data, $d$, we are interested in deriving ${\widetilde P}(\delta \mu_{3n}|d, \, \mathcal{H})$, the posterior probability distribution on $\delta \mu_{3n}$, for a flat prior on $\delta\mu_{3n}$ (here $\mathcal{H}$ denotes the hypothesis, which is the parametric model we employ.). Towards this, we use Eqs.~(\ref{eq:bound_deltamu30}),~(\ref{eq:bound_deltamu32}), and (\ref{eq:bound_deltamu33}), along with the two-dimensional posterior distribution $P(\delta \hat{\phi_b}, \, \nu|d, \, \mathcal{H})$, for different GW events. For example, to derive ${\widetilde P}(\delta \mu_{30}|d, \, \mathcal{H})$ we will use Eq.~(\ref{eq:bound_deltamu30}) and $P(\delta \hat{\phi_2}, \, \nu|d, \, \mathcal{H})$.
The probability distribution $P(\delta \hat{\phi_b}, \, \nu|d, \, \mathcal{H})$ is computed for flat priors on $\delta \hat{\phi_b}$ and mass ratio. Therefore in the Bayesian framework the samples of $\delta \mu_{3n}$, derived from $P(\delta \hat{\phi_b}, \, \nu|d, \, \mathcal{H})$, need to be \emph{reweighted} to obtain posterior $\widetilde{P}(\delta \mu_{3n}|d, \, \mathcal{H})$ that assume \emph{flat priors} on $\delta \mu_{3n}$ as
\begin{equation}
\begin{split}
    \widetilde{P}(\delta \mu_{3n}|d, \, \mathcal{H}) \propto \bigg[ \int d \nu \, d(\delta \hat{\phi}_{b}) P(\delta \mu_{3n}|\delta \hat {\phi_b}, \, \nu, \, \mathcal{H})\; \\
    {\widetilde P}(\delta \hat{\phi_b},\, \nu|d, \, \, \mathcal{H})\;\bigg] \times
    \frac{{\widetilde \Pi}(\delta \mu_{3n} | \mathcal{H}) }{\Pi (\delta \mu_{3n}|\mathcal{H})} \, . 
    \label{eq:Pdmu3}
\end{split}    
\end{equation}

In the above equation, the tilde denotes flat priors or posteriors derived assuming flat prior on the corresponding parameters. 
Hence ${\widetilde \Pi}(\delta \mu_{3n}|\mathcal{H})$ denotes flat prior on $\delta \mu_{3n}$ and ${\widetilde P}(\delta \hat{\phi_b}, \, \nu|d, \, \mathcal{H})$ denote posterior assuming flat priors on $\delta \hat {\phi_b}$ and mass ratio. The prior distribution, ${\widetilde \Pi}(\delta \mu_{3n} | \mathcal{H})$, is chosen to be uniform between [-30, 30]. While $P(\delta \mu_{3n}|\delta \hat {\phi_b}, \, \nu, \, \mathcal{H})$ takes care of the coordinate transformation between $(\delta \hat {\phi_b}, \, \nu)$ to $\delta \mu_{3n}$ [using Eqs.~(\ref{eq:bound_deltamu30})--(\ref{eq:bound_deltamu33})], $\Pi(\delta \mu_{3n} | \mathcal{H})$ in the above equation is simply $P(\delta \mu_{3n}|\delta \hat{\phi_b}, \, \nu, \, \mathcal{H})$ for the flat prior on $\delta \hat {\phi_b}$ and mass ratio. A detailed derivation of the above equation is provided in Sec.~\ref{appendixbayes} of the Supplemental Material.

As GW190814 and GW190412 are the only two events for which a confident detection of higher modes was possible, we will restrict to these two events for our purposes. We use the parameter estimation samples for $\delta \hat{\phi}_{b}$ and symmetric mass ratio ($\nu$) from the  \emph{GWTC-3 Data Release}~\cite{lvc:dataGWTC3} for these two events, analyzed with different waveform approximants, and estimate the bounds on the parameters $\delta \mu_{3n}$ following this procedure.

While estimating the posterior probability distribution of $\delta \mu_{30}$, we need to ensure that the values of $\delta \mu_{30}$ obtained from a pair of $(\delta{\hat\phi_2}, \, \nu)$ should be real by imposing the condition [see Eq.~(\ref{eq:bound_deltamu30})],
\begin{equation}\label{eq:condition}
  \delta \hat{\phi}_{2} \leq \frac{K_{1}(\nu)}{\phi_{2}^{\rm GR}}.
\end{equation}
In order to realize this, we discard those samples of $\delta \mu_{30}$ which do not meet the above condition. If we need to remove relatively large number of samples, that means the event is uninformative. We find that for events except GW190814 and GW190412, majority of the samples do not meet this condition. This simply is a reflection of the fact that we are trying to test deviation in octupole moment when its presence is barely there in the signal.

\section{Results and conclusions}\label{sec:results}
The posteriors of the leading and two subleading octupole deformation parameters $\delta \mu_{30}$, $\delta \mu_{32}$, and $\delta \mu_{33}$ for GW190412 and GW190814 obtained by the above-mentioned procedure are shown in Fig.~\ref{fig:dmu30-dmu32-dmu33-bounds}. Among the detected events, GW190814 provides the tightest constraints on all the octupole parameters. This is expected as GW190814 is the most unequal mass binary (mass ratio, $q=0.112^{+0.008}_{-0.009}$) among the GW events in the GWTC-3 and asymmetric systems get stronger contributions from non-quadrupolar moments.

In addition to the individual event analysis in Fig.~\ref{fig:dmu30-dmu32-dmu33-bounds}, we have also obtained the combined bounds on $\delta \mu_{3n}$ using data from multiple events under the assumption that the same value of $\delta \mu_{3n}$ is shared across all the events.
The joint constraints on these parameters are obtained by multiplying the individual likelihoods from the events, GW190412 and GW190814, analyzed with {\tt Phenom} and {\tt SEOB} waveforms. In the joint analysis the most tightly constrained parameter is $\delta \mu_{30}$ and the most weakly constrained parameter is $\delta \mu_{33}$. The posterior on higher-order mass octupole deformation parameters, such as $\delta \mu_{34}$, are mostly uninformative and not shown here. Detections of unequal mass binaries in the future with a larger signal-to-noise ratio will enable us to probe higher PN pieces in the mass octupole moment.

Bounds from the two different kinds of waveform approximants show excellent agreement with each other. On all occasions the posterior distributions on $\delta \mu_{3n}$ are statistically consistent with $\delta \mu_{3n}=0$ within 90\% credible interval.
This is the first reported bound on the different PN pieces in mass-type octupole moment of compact binary complementing the previous consistency tests in Refs.~\cite{Capano:2020dix,Puecher:2022sfm}. It is interesting that the bounds on $\delta \mu_{33}$ also confirm the consistency of the octupolar tail radiation with the predictions of GR.

The posterior distributions on current quadrupole deformation parameters are also largely uninformative and not reported here. In the future, the detections of high mass ratio and highly spinning binaries with larger signal-to-noise ratio will enhance the contribution of the current quadrupole to the flux making its measurement with good precision possible.

Last, it is instructive to ask if the GR violations in the $\delta \hat\phi_b$ posteriors can be captured by the mapping proposed in this work.
This is examined in Fig.~\ref{fig:dmu30-dmu32-dmu33-gr-nongr} of the Supplemental Material (see the texts in Sec.~\ref{appendixgrnongr} of the Supplemental Material for more details). We consider a GW190814-like system and simulate GR violations with different $\sigma$ values in the $\delta \hat\phi_b$ posteriors. We find that the derived posteriors on $\delta \mu_{3n}$, through this mapping, will be able to detect deviations at different $\sigma$ values in the $\delta \hat\phi_b$ posteriors from GR.

To conclude, the parametrized multipolar waveforms could play a pivotal role in testing GR with current and next-generation GW detectors. The next-generation GW detectors will observe more diverse classes of compact binaries, thereby allowing us to probe even higher multipoles of compact binaries, and the parametrization introduced here will be crucial in such scenarios. Development of an infrastructure that can directly sample over the multipole parameters is planned for future work which should be able to probe the multipole structure without relying on the approximate mapping we have invoked here. The direct inference of multipole parameters from the GW data could provide even more stringent constraints as one will potentially gain information from multiple PN coefficients in the phase.
Moreover, the inclusion of multipole parameters in the amplitude will likely play a critical role in this framework.

\acknowledgments
This material is based upon work supported by the NSF’s LIGO Laboratory, which is a major facility fully funded by the National Science Foundation (NSF). The author is grateful for computational resources provided by the LIGO Laboratory and supported by National Science Foundation Grants No. PHY-0757058 and No. PHY-0823459.
It is a pleasure to thank K. G. Arun, who suggested this problem to me, encouraged me to pursue it, and for comments on the draft. The author is grateful to A. Gupta, S. Kastha, and B. S. Sathyaprakash for invaluable discussions and/or comments on the manuscript. We thank M. Saleem for critical reading of the manuscript and providing useful comments. We also thank P. Saini, S. A. Bhat, and P. D. Roy for comments on the manuscript. We are thankful to N. J. McDaniel, A. Laddha, and S. Datta for valuable discussions. P.M. acknowledge the support of the Core Research Grant No. CRG/2021/004565 of the Science and Engineering Research Board of India and a grant from the Infosys foundation.
This research has made use of data obtained from the Gravitational Wave Open Science Center (www.gw-openscience.org), a service of LIGO Laboratory, the LIGO Scientific Collaboration and the Virgo Collaboration. Virgo is funded by the French Centre National de Recherche Scientifique (CNRS), the Italian Istituto Nazionale della Fisica Nucleare (INFN), and the Dutch Nikhef, with contributions by Polish and Hungarian institutes. This manuscript has the LIGO preprint number {\tt P2300150}.
Analyses in this paper made use of NumPy~\cite{Numpy2011}, SciPy~\cite{Scipy2020}, IPython~\cite{IPython2007}, Matplotlib~\cite{Matplotlib2007}, Corner~\cite{corner}, Jupyter~\cite{Jupyter}, Seaborn~\cite{Seaborn} software packages.


\bibliographystyle{apsrev}
\bibliography{ref-list}

\clearpage

\onecolumngrid
\appendix*

\section{Supplemental Materials}

\subsection{Tracking different PN pieces of the mass type octupole moment}\label{appendixoct}
With the parametrization introduced in Eq.~(\ref{eq:parametrized-Uijk}) of the main paper, the different PN contributions from the mass type radiative octupole moment to the GW flux, and hence to the GW phasing, can be separately kept track of. To derive such a parametrized octupolar GW phasing, we follow the MPM formalism~\cite{BDI95} along the lines of Refs.~\cite{Kastha:2018bcr,Kastha:2019}. The parametrized multipolar flux schematically reads as

\begin{align}\label{eq:Flux}
    \mathcal{F} = \frac{32}{5} \nu^{2} x^{5} \mu_{2}^{2} \bigg[1 + \mathcal{F}_{\rm 1 PN} \,  x + \mathcal{F}_{\rm 1.5 PN} \,  x^{3/2} + \mathcal{F}_{\rm 2 PN} \,  x^{2} + \mathcal{F}_{\rm 2.5 PN} \,  x^{5/2} + \mathcal{O}(x
^3) \bigg] \, ,
\end{align}
where $x=(GM\omega/c^{3})^{2/3}$ is a PN parameter, $\omega$ the orbital angular frequency of the binary.

The leading order contribution from the mass-type octupole moment ($U_{ijk}$) to the GW flux first appears at 1PN:
\begin{align}
    {\mathcal{F}}_{\rm 1 PN}^{\rm Oct} =  \left(\frac{\mu_{30}}{\mu_{2}}\right)^{2} \left[ \frac{1367}{1008} - \frac{1367}{252} \nu  \right] \, .
\end{align}
Therefore the leading order contribution from $U_{ijk}$ to the GW phasing appears at 1PN and is given by Eq.~(\ref{eq:phi2}) in the main paper.

The next PN contribution (i.e., 1PN correction to $U_{ijk}$; note that there is no 0.5PN correction to $U_{ijk}$.) from $U_{ijk}$ to the GW flux makes appearance at 2PN which reads as
\begin{align}
   &{\mathcal{F}}_{\rm 2 PN}^{\rm Oct} =  \left[ -\frac{1367 \, \mu_{30}^2}{168 \, \mu_{2}^2}-\frac{8201 \, \mu_{30} \, \mu_{32}}{3024 \, \mu_{2}^2} 
    + \left( \frac{17771 \, \mu_{30}^2}{504 \, \mu_{2}^2}+\frac{1139 \, \mu_{30} \, \mu_{32}}{84 \, \mu_{2}^2} \right) \nu 
     - \left(\frac{1367 \, \mu_{30}^2}{126 \, \mu_{2}^2} + \frac{2050 \, \mu_{30} \, \mu_{32}}{189 \, \mu_{2}^2} \right) \nu^{2}\right] \, .
\end{align}
Thus the 1PN correction to the octupole moment first appears at 2PN in the phasing. The 2PN phasing coefficient reads
\begin{align}
\phi_{4} &= \Bigg(\frac{242245}{5292}+\frac{4525}{5292} \nu +\frac{145445}{5292} \nu ^2 +\left(\frac{\mu_{30}}{\mu_{2}}\right)^{2} \Bigg[-\frac{772355}{21168} +\frac{580975}{3024} \nu -\frac{977405}{5292} \nu ^2 \Bigg] + \left(\frac{\mu_{30} \, \mu_{32}}{\mu_{2}^{2}}\right) \Bigg[\frac{41005}{1512} \nonumber \\
&\quad -\frac{5695}{42} \nu + \frac{20500}{189} \nu ^2 \Bigg] 
+\left(\frac{\mu_{30}}{\mu_{2}}\right)^{2}  \left(\frac{\epsilon_{2}}{\mu_{2}}\right)^{2}  \Bigg[\frac{6835}{9072} -\frac{6835}{1134} \nu +\frac{6835 \nu ^2}{567}\Bigg]
 +\left(\frac{\mu_{30}}{\mu_{2}}\right)^{4}\Bigg[\frac{9343445}{508032}-\frac{9343445}{63504} \nu \nonumber \\
&\quad+\frac{9343445}{31752} \nu ^2\Bigg] 
+\left(\frac{\mu_{4}}{\mu_{2}}\right)^{2}\Bigg[-\frac{89650}{3969}+\frac{179300}{1323} \nu  - \frac{89650}{441}\nu^2\Bigg] 
+\left(\frac{\epsilon_{2}}{\mu_{2}}\right)^{2} \Bigg[-\frac{785}{378}+\frac{7115}{756}\nu -\frac{835}{189} \nu ^2\Bigg] \nonumber \\
&\quad+\left(\frac{\epsilon_{2}}{\mu_{2}}\right)^{4}\Bigg[\frac{5}{648}-\frac{5 }{81} \nu+\frac{10}{81} \nu ^2
 \Bigg] + \left(\frac{\epsilon_{3}}{\mu_{2}}\right)^{2}\Bigg[-\frac{50}{63}+\frac{100}{21} \nu-\frac{50}{7} \nu ^2\Bigg]
 \Bigg) \, .
\end{align}
With the assumptions $\mu_{l;l\neq3}=\epsilon_{l}=1$ and $\mu_{3 n^{\prime};n^{\prime}\neq 2}=1$, the 2PN phasing coefficient reduces to
\begin{align}\label{eq:phi4dmu32}
  &\phi_{4} = \phi_{4}^{\rm GR} +  K_{3}(\nu) \, \big(\mu_{32}-1\big),
\end{align}
where $\phi_{4}^{\rm GR}$ and $K_{3}(\nu)$ are given by 
\begin{align}
    &\phi_{4}^{\rm GR} = \left( \frac{15293365}{508032}+\frac{27145}{504} \nu + \frac{3085}{72} \nu^2 \right) \, ,\label{eq:phi4GR} \\
    &K_{3}(\nu) = \left( \frac{41005}{1512} -\frac{5695}{42} \nu + \frac{20500}{189} \nu ^2 \right) \, . \label{eq:K3}
\end{align}
Comparing the Eq.~(\ref{eq:phi4dmu32}) with the parametrization, $\phi_{4} \rightarrow \phi_{4}^{\rm GR} (1+\delta \hat{\phi}_{4})$, we have:
\begin{align}
    \delta \hat{\phi}_{4} = \frac{K_{3}(\nu) \, \left(\mu_{32}-1\right)}{\phi_{4}^{\rm GR}}\, .
\end{align}
Further, substituting $\mu_{32}\rightarrow(1+\delta \mu_{32})$ into the above equation we got the following expression for $\delta \mu_{32}$:
\begin{align}\label{eq:dmu32}
    \delta \mu_{32} = \frac{\delta \hat{\phi}_{4} \, \phi_{4}^{\rm GR}}{K_{3}(\nu)} \, .
\end{align}
The 1.5PN correction to $U_{ijk}$, that is the octupolar tail, first appears at 2.5PN in the GW flux:
\begin{align}
& {\mathcal{F}}_{\rm 2.5PN}^{\rm Oct} = \pi \, \left(\frac{\mu_{30} \, \mu_{33}}{\mu_{2}^{2}}\right) \left(\frac{16403}{2016}-\frac{16403}{504} \nu \right)\,.
\end{align}
Hence the tail contribution from $U_{ijk}$ to the GW phase shows up at the 2.5PN-logarithmic term. The 2.5PN logarithmic phasing coefficient reads,
\begin{align}
\phi_{5l} &= \pi \Bigg( \frac{12080}{63} -\frac{3680}{21} \nu +
\left(\frac{\mu_{30}}{\mu_{2}}\right)^{2} \Bigg[ - \frac{27340}{189} + \frac{109360}{189} \nu \Bigg] + \left(\frac{\mu_{30} \, \mu_{33}}{\mu_{2}^{2}}\right) \Bigg[\frac{82015}{756} - \frac{82015}{189} \nu \Bigg] \nonumber \\
&\quad+ \left(\frac{\epsilon_{2}}{\mu_{2}}\right)^{2} \Bigg[ -\frac{20}{9} + \frac{80}{9} \nu \Bigg] \Bigg) \, .
\end{align}
Assuming $\mu_{l;l\neq3}=\epsilon_{l}=1$ and $\mu_{3 n^{\prime};n^{\prime}\neq 3}=1$, the 2.5PN phasing coefficient simplifies to,
\begin{align}\label{eq:phi5ldmu33}
    &\phi_{5l} = \phi_{5l}^{\rm GR} + K_{4}(\nu) \, \big(\mu_{33}-1\big),
\end{align}
where $\phi_{5l}^{\rm GR}$ and $K_{4}(\nu)$ are given by 
\begin{align}
    &\phi_{5l}^{\rm GR} = \pi \left( \frac{38645}{252}-\frac{65}{3} \nu  \right) \, ,\label{eq:phi5lGR}\\
    &K_{4}(\nu) = \pi \left( \frac{82015 }{756}-\frac{82015}{189} \nu  \right)\, \label{eq:K4}.
\end{align}
Again comparing the above equation (Eq.~\ref{eq:phi5ldmu33}) with the parametrization, $\phi_{5l} \rightarrow \phi_{5l}^{\rm GR} (1+\delta \hat{\phi}_{5l})$ and replacing $\mu_{33}$ with $(1+\delta \mu_{33})$, we obtained the following expression for $\delta \mu_{33}$:
\begin{align}\label{eq:dmu33}
    \delta \mu_{33} = \frac{\delta \hat{\phi}_{5l} \, \phi_{5l}^{\rm GR}}{K_{4}(\nu)} \, .
\end{align}

\subsection{Analysis framework}\label{appendixbayes}
Within the framework of Bayesian inference, measuring the parameter $\delta \mu_{3n}$ amounts to obtaining the posterior probability density function $P(\delta \mu_{3n} | d, \mathcal{H})$, with $d$ denotes the detector data and $\mathcal{H}$ denotes the model. Using Bayes' theorem,
\begin{equation}\label{eq:Bayes}
    P(\delta \mu_{3n} | d, \mathcal{H})  =  \frac{P(\delta \mu_{3n} | \mathcal{H}) \, P(d | \delta \mu_{3n}, \mathcal{H} )}{P(d | \mathcal{H})},
\end{equation}
where, $P (\delta \mu_{3n} | \mathcal{H})$ is the prior probability density function, $P(d | \delta \mu_{3n}, \mathcal{H} )$ is the likelihood function, and  $P(d | \mathcal{H})$ [with $P(d | \mathcal{H})= \int d(\delta \mu_{3n}) P (\delta \mu_{3n} | \mathcal{H}) \, P(d | \delta \mu_{3n}, \mathcal{H} )$] is the evidence. 
The above equation with uniform prior on $\delta \mu_{3n}$ (i.e., $P(\delta \mu_{3n} | \mathcal{H})={\widetilde \Pi}(\delta \mu_{3n} | \mathcal{H})$) can be written as follows,
\begin{align} \label{eq:mapping}
\widetilde{P}(\delta \mu_{3n} | d, \mathcal{H}) & =  \frac{{\widetilde \Pi}(\delta \mu_{3n} | \mathcal{H}) \, P(d | \delta \mu_{3n} , \mathcal{H})}{\widetilde{P}(d | \mathcal{H})} \nonumber\\
 & = \frac{{\widetilde \Pi}(\delta \mu_{3n} | \mathcal{H}) \, \int d \nu \, d(\delta \hat{\phi}_{b}) \, P(d|\delta \hat{\phi}_b, \nu, \mathcal{H}) \, P(\delta \hat{\phi}_b, \nu | \delta \mu_{3n}, \mathcal{H})}{\widetilde{P}(d | \mathcal{H})} \nonumber\\
 & = \frac{{\widetilde \Pi}(\delta \mu_{3n} | \mathcal{H})}{\widetilde{P}(d | \mathcal{H})} \, \times \, \int d \nu \, d(\delta \hat{\phi}_{b}) P(d|\delta \hat{\phi}_b, \nu, \mathcal{H}) \, \underbrace{\frac{{\widetilde \Pi}(\delta \hat{\phi}_b, \nu | \mathcal{H}) \, P(\delta \mu_{3n} | \delta \hat{\phi}_b, \nu, \mathcal{H})}{\Pi (\delta \mu_{3n}| \mathcal{H})}}_{P(\delta \hat{\phi}_b, \nu | \delta \mu_{3n}, \mathcal{H})} \, , \nonumber\\
\end{align}
where, $\widetilde{P}(d | \mathcal{H})$ is the evidence for the uniform prior on $\delta \mu_{3n}$ [i.e., $\widetilde{P}(d | \mathcal{H}) \equiv \int d(\delta \mu_{3n}) {\widetilde \Pi}(\delta \mu_{3n} | \mathcal{H}) \, P(d | \delta \mu_{3n}, \mathcal{H} )$], $P(d|\delta \hat{\phi}_b, \nu, \mathcal{H})$ is the likelihood function of the gravitational wave data given the parameters $\delta \hat{\phi}_b$ and $\nu$, $P(\delta \mu_{3n} | \delta \hat{\phi}_b, \nu, \mathcal{H})$ takes care of the coordinate transformation between $\{\delta \hat{\phi}_{b}, \, \nu\}$ and $\delta \mu_{3n}$, ${\widetilde \Pi}(\delta \hat{\phi}_b, \nu | \mathcal{H})$ is the flat prior on $\{\delta \hat{\phi}_{b}, \, \nu\}$, and $\Pi (\delta \mu_{3n}| \mathcal{H})$ is given by,
\begin{equation}\label{eq:old-prior}
    \Pi (\delta \mu_{3n} | \mathcal{H}) \equiv \int d \nu \, d(\delta \hat{\phi}_{b}) \, {\widetilde \Pi}(\delta \hat{\phi}_b, \nu | \mathcal{H}) \, P(\delta \mu_{3n} | \delta \hat{\phi}_b, \nu, \mathcal{H}) \, .
\end{equation}
Therefore, $\Pi(\delta \mu_{3n} | \mathcal{H})$ is simply the distribution of $\delta \mu_{3n}$ derived from the uniform prior on $\{\delta \hat{\phi}_{b}, \, \nu\}$ and the relation between $\delta \mu_{3n}$ and $\{\delta \hat{\phi}_{b}, \, \nu\}$. The Eq.~\ref{eq:mapping} can be further simplified as
\begin{align} \label{eq:reweight}
\widetilde{P}(\delta \mu_{3n} | d, \mathcal{H}) & = {\widetilde \Pi}(\delta \mu_{3n} | \mathcal{H}) \, \times \frac{{\widetilde P}_{IT}(d|\mathcal{H})}{{\widetilde P}(d|\mathcal{H})} \, \times \, \int d \nu \, d(\delta \hat{\phi}_{b}) \bigg[ \underbrace{\frac{{\widetilde \Pi}(\delta \hat{\phi}_b, \nu | \mathcal{H}) \, P(d|\delta \hat{\phi}_b, \nu, \mathcal{H})}{{\widetilde P}_{IT}(d|\mathcal{H})}}_{\widetilde{P}(\delta \hat{\phi}_b, \nu | d, \mathcal{H})} \, \times \, \frac{P(\delta \mu_{3n} | \delta \hat{\phi}_b, \nu, \mathcal{H})}{\Pi (\delta \mu_{3n}|\mathcal{H})} \bigg] \nonumber\\
 & = \bigg[ \int d \nu \, d(\delta \hat{\phi}_{b}) P(\delta \mu_{3n}|\delta \hat {\phi_b}, \, \nu, \, \mathcal{H})\; {\widetilde P}(\delta \hat{\phi_b},\, \nu|d, \, \, \mathcal{H})\;\bigg] \times \frac{{\widetilde \Pi}(\delta \mu_{3n} | \mathcal{H}) }{\Pi (\delta \mu_{3n}|\mathcal{H})} \times \frac{{\widetilde P}_{IT}(d|\mathcal{H})}{{\widetilde P}(d|\mathcal{H})} \, ,
\end{align}
where, $\widetilde{P}(\delta \hat{\phi}_b, \nu | d, \mathcal{H})$ is the posterior probability density of $\{\delta \hat{\phi}_{b}, \, \nu\}$ and ${\widetilde P}_{IT}(d|\mathcal{H})$ [with ${\widetilde P}_{IT}(d|\mathcal{H}) = \int d \nu \, d(\delta \hat{\phi}_{b}) \, {\widetilde \Pi}(\delta \hat{\phi}_b, \nu | \mathcal{H}) \, P(d|\delta \hat{\phi}_b, \nu, \mathcal{H})$] is the corresponding evidence derived assuming a uniform prior on $\{\delta \hat{\phi}_{b}, \, \nu\}$. The numerical factor $\frac{{\widetilde P}_{IT}(d|\mathcal{H})}{{\widetilde P}(d|\mathcal{H})}$ is an overall normalization constant in the above equation and can be ignored if one is only interested in estimating $\widetilde{P}(\delta \mu_{3n} | d, \mathcal{H})$. 

The probability density function $P(\delta \mu_{3n} | \delta \hat{\phi}_b, \nu, \mathcal{H})$ maps the posterior samples of \{$\delta \hat{\phi}_{b},  \nu$\} to samples of $\delta \mu_{3n}$ via the relations given by Eq.~(\ref{eq:bound_deltamu30}) in the main paper for $\{\delta \hat{\phi}_{2}, \nu\} \rightarrow \{ \delta  \mu_{30}\}$, Eq.~(\ref{eq:dmu32}) for $\{\delta \hat{\phi}_{4}, \nu\} \rightarrow \{ \delta  \mu_{32}\}$ and Eq.~(\ref{eq:dmu33}) for $\{\delta \hat{\phi}_{5l}, \nu\} \rightarrow \{ \delta  \mu_{33}\}$. Moreover, $\delta \mu_{3n}$ is a function $f_{3n}(\delta \hat{\phi}_{b},\nu)$ of $\delta \hat{\phi}_{b}$ and $\nu$; hence given a value of the pair \{$\delta \hat{\phi}_{b},  \nu$\}, $\delta \mu_{3n}$ is uniquely determined using the relations mentioned above. Therefore, $P(\delta \mu_{3n} | \delta \hat{\phi}_b, \nu, \mathcal{H})$ becomes delta function:
\begin{align}
    P(\delta \mu_{3n} | \delta \hat{\phi}_b, \nu, \mathcal{H}) = \delta(\delta \mu_{3n} - f_{3n}(\delta \hat{\phi}_{b},\nu))\, .
\end{align}
In practice one needs to take the posterior samples of \{$\delta \hat{\phi}_{b},  \nu$\} for different GW events and to estimate $\delta \mu_{3n}$ for each sample through the relations mentioned above; and then reweight these samples by the probability $\frac{{\widetilde \Pi}(\delta \mu_{3n} | \mathcal{H}) }{\Pi (\delta \mu_{3n}|\mathcal{H})}$.

\begin{figure*}
    \centering
    \includegraphics[width=1.0\textwidth]{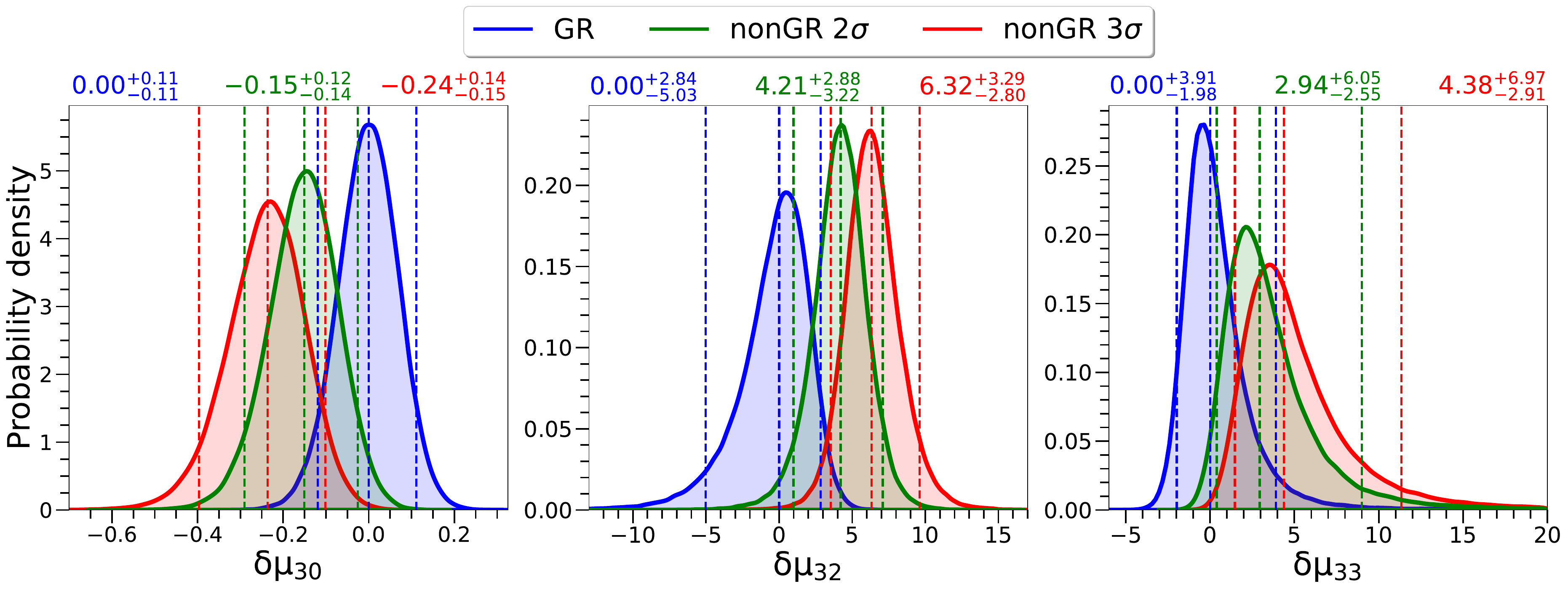}
    \caption{The probability distributions of $\delta \mu_{3n}$ for ``GR", ``nonGR 2$\sigma$" and ``nonGR 3$\sigma$" cases (see the texts in Sec.~\ref{appendixgrnongr} of the Supplemental Material for more details). The widths of the probability distributions of $\{\nu, \delta \hat{\phi}_{b}\}$ are chosen similar to GW190814. In the ``nonGR 2$\sigma$" and ``nonGR 3$\sigma$" cases the means of the $\delta \hat{\phi}_{b}$-distributions are 2$\sigma$ and 3$\sigma$ away from zero respectively. The colored vertical dashed lines indicate the 90\% confidence intervals and median values. The probability distributions of $\delta \mu_{3n}$ are peaking at zero in the ``GR" cases while in the ``nonGR 2$\sigma$" and ``nonGR 3$\sigma$" cases the distributions of $\delta \mu_{3n}$ are excluding zero at 90\% credibility.}
    \label{fig:dmu30-dmu32-dmu33-gr-nongr}
\end{figure*}

\subsection{Detecting GR violation}\label{appendixgrnongr}
Here we investigate to what extent deviations in the phase deformation coefficients ($\delta \hat{\phi}_{b}$) will be propagated to the probability distributions of the octupole deformation parameters ($\delta \mu_{3n}$) through this mapping. For demonstration, we consider a GW190814-like system. In order to simulate the GR case, we have drawn samples for $\{\nu, \delta \hat{\phi}_{2} \}$ from a bivariate normal distribution with means, $\{0.09, 0.0\}$, standard deviations, $\{0.01, 0.05 \}$, and correlation coefficient, $-0.29$. After generating the samples for $\{\nu, \delta \hat{\phi}_{2} \}$, we estimate the probability distribution of $\delta \mu_{30}$ (shown in the blue curve in the left-most panel of Fig.~\ref{fig:dmu30-dmu32-dmu33-gr-nongr}) following the above-mentioned procedure. As expected, we see that the probability distribution of $\delta \mu_{30}$ is sharply peaking at zero (blue curve in the left-most panel of Fig.~\ref{fig:dmu30-dmu32-dmu33-gr-nongr}). In the second scenario, which models a 2$\sigma$ violation of GR, we have chosen the mean associated with $\delta \hat{\phi}_{2}$ to be 0.1 (2$\sigma$ deviation from zero i.e., GR value) and kept all other parameters of the bivariate normal distribution at the same values as before. Again we obtain the probability distribution of $\delta \mu_{30}$ (shown in the green curve in the left-most panel of Fig.~\ref{fig:dmu30-dmu32-dmu33-gr-nongr}) and find that the distribution of $\delta \mu_{30}$ excludes zero (GR value) at 90\% credibility. 
To simulate another non-GR scenario, a GR violation at 3$\sigma$, we consider the mean of $\delta \hat{\phi}_{2}$ to be 0.15 (3$\sigma$ deviation from zero i.e., GR value) and obtain the probability distribution of $\delta \mu_{30}$ (red curve in the left-most panel of Fig.~\ref{fig:dmu30-dmu32-dmu33-gr-nongr}).

We also performed the similar exercises and obtain the probability distributions on $\delta \mu_{32}$ (middle panel of Fig.~\ref{fig:dmu30-dmu32-dmu33-gr-nongr}) and $\delta \mu_{33}$ (rightmost panel of Fig.~\ref{fig:dmu30-dmu32-dmu33-gr-nongr}). We find that in the ``GR" cases the probability distributions of \{$\delta \mu_{3n}$\} are peaking at zero (or, very close to zero; the median values of the distributions are zero) while in the ``nonGR 2$\sigma$" and ``nonGR 3$\sigma$" cases the distributions of \{$\delta \mu_{3n}$\} are excluding zero at 90\% credibility. This demonstrates that the proposed mapping is successful in mapping the posteriors of the PN deformations reliably to the posteriors of the corresponding octupole deformation parameters.

\end{document}